# Navigation through the complex world – the neurophysiology of decision-making processes

Ugurcan Mugan[1], Seiichiro Amemiya[2], Paul S. Regier[3], A. David Redish[1*]
1 Department of Neuroscience, University of Minnesota, United States
2 RIKEN Center for Brain Science, Saitama, Japan
3 Department of Psychiatry, University of Pennsylvania, United States
* Corresponding author: redish@umn.edu

## Abstract
Current theories suggest that adaptive decision-making necessitates the interaction between multiple decision-making systems. The computational definitions of different models of decision-making suggest interactions with task demands and complexity. We review these computational theories and derive experimental predictions that will shed light on the underlying neurobiological mechanisms. We use a well-established multi-strategy task and novel neurophysiological analyses from hippocampus and striatum as a case study in the interaction between task structure and navigational complexity. This approach reveals how task structure and navigational complexity interact with each other to identify differences between habitual and planned action choices.

## Multiple strategies for behavioral control
Studies of decision-making suggest multiple competing behavioral control systems, two of which are prospective (planning, deliberative, goal-directed) and procedural (habitual). Current theories suggest that these two systems depend on different neural circuits, are optimized for different situations and contexts, and are reflected in subtle behavioral differences during decision-making (Hull 1943, O'Keefe and Nadel 1978, Balleine and Dickinson 1998, Redish 1999, Daw, Niv et al. 2005, Botvinick, Niv et al. 2009, Redish 2013). Prospective algorithms are mathematically similar to action-outcome proposals, as well as cognitive map navigation, both of which depend on expectations of the future (Averbeck and O'Doherty 2022, Friedman and Robbins 2022). Prospective systems depend on learning the structure of the world and the consequences of the actions within it, which can then be used for planning (Miller, Botvinick et al. 2017). As such, prospective behaviors can be flexible (even entirely novel) (Gupta, van der Meer et al. 2010, Pfeiffer and Foster 2013, Redish 2016), arising from evaluations of internal representations of imagined outcomes (Buckner and Carroll 2007, Hassabis, Kumaran et al. 2007, Johnson and Redish 2007). In contrast, procedural algorithms are mathematically similar to route navigation, implicit skill learning, and stimulus-response (or situation-action) relationships (Jog, Kubota et al. 1999, Yin and Knowlton 2006, Graybiel 2008). In procedural decision processes, the most appropriate cached actions in a given situation are learned through retrospective evaluation of experienced rewards and failures (Barnes, Kubota et al. 2005, Smith and Graybiel 2016). Therefore, procedural control enables computationally efficient decision making, but at the cost of behavioral flexibility.

Current theories also include a third decision-making algorithm usually referred to as Pavlovian action-selection, in which a limited repertoire of actions can be learned and released in response to associated cues (Breland and Breland 1961). Actions governed by this behavioral controller are species-specific and species-important responses. Since these actions are innate responses to environmental stimuli, the release of Pavlovian actions are fast and computationally inexpensive (Dayan, Niv et al. 2006). Pavlovian repertoires such as grooming sequences and nest-building behavior in rodents and birds are developed in evolutionary time, as are their initial unconditioned signaling stimuli such as the stimulation of an itch initiating grooming or environmental cues and/or hormonal changes associated with nesting. However, these sequences can be generalized and released in other contexts and in response to other stimuli that may be learned or experiences during the lifetime of the animal (e.g., classical Pavlovian conditioning). Since responses are instinctual, the Pavlovian system can be engaged even in entirely novel situations (Huys, Eshel et al. 2012, Lally, Huys et al. 2017, Mobbs, Headley et al. 2020). Given that the tasks that we will consider require complex navigation strategies and are non-threatening and non-aversive, we will consider the Pavlovian system outside the scope of this review and will not be detailing its role in adaptive decision-making.

While classical decision-theories dichotomize prospective and habit-based decision-making processes, more recent theories have proposed intermediate and hybrid variations, which contain aspects of both prospection and habit. Some of these intermediate variants rely on using the state transition information and the stability of those state transitions to find reliable action chains that can be collapsed and represented as a single action sequence.  (Dayan 1993, Gershman 2018); others propose that action chains, chosen initially by prospection, are separately cached based on their expected values, such that they span intermediary subgoals (Graybiel 1998, Botvinick 2008, Dezfouli, Lingawi et al. 2014, Cushman and Morris 2015). These hybrid theories suggest that behavioral control can further intersperse prospective and procedural processes across these subgoals — for example, prospection over subgoals with cached action chains between them.

Past research has emphasized the competition between procedural (habit, based on situational release of action-chains) and prospective (planned, based on explicit goal-directed processes) control of behavior and suggested that they arise based on the animal's experience and the consistency of the environment (Jog, Kubota et al. 1999, Smith and Graybiel 2016, Robbins and Costa 2017). When the optimal behavior is relatively consistent across time, habitual control is favored, particularly after extensive experience. In contrast, planning is favored when the optimal policy remains variable and in conditions of limited experience (Daw, Niv et al. 2005, Rich and Shapiro 2009, Durstewitz, Vittoz et al. 2010, Karlsson, Tervo et al. 2012, Powell and Redish 2014). Importantly, however, planning requires that an animal has an adequate understanding of the process of their task.

Classically, three tests have been used to measure the competition between the two systems: 1) behavioral outcome-devaluation (Balleine, Garner et al. 1995, Balleine and Dickinson 2005), 2) changing action-reward probabilities (Rushworth and Behrens 2008, Chen, Knep et al. 2021, Wilson, Bonawitz et al. 2021), and 3) changing the consequences of actions (Daw, O'Doherty et al. 2006, Amemiya and Redish 2016, Hasz and Redish 2018). These three tests aim to elucidate how changing the reward value, reward probability, and/or action-reward pairing influences how behavioral control shifts between the two systems (i.e., arbitration between habit and planning). All three tests rely on a theoretical model of dichotomy between different decision-making systems. However, they are used to probe different aspects of changes to the decision-making strategy due to outcome changes. In general, studies have found that, initially, behavioral control is largely prospective, but that it becomes increasingly habitual (stimulus-response) with continued training (O'Keefe and Nadel 1978, van der Meer, Johnson et al. 2010, Redish 2016).

Outcome devaluation studies approach this question by making previously appetitive rewards less valuable. Classical experiments of outcome devaluation devalue positive food rewards by associating the reward with an aversive lithium chloride injection (Balleine, Garner et al. 1995). Typically, after devaluation, the animal is no longer motivated to achieve the reward. However, these studies have shown that despite reward devaluation, extensively-trained animals often continue to take the action that previously led to reward. This behavioral inflexibility is hypothesized to be a consequence of the transition to procedural control. At the level of the brain, studies have shown that habit-based decision making is governed, in part, by the dorsolateral striatum. For example, in studies that disrupt dorsolateral striatal function, actions remain sensitive to devaluation (Yin, Knowlton et al. 2004, Yin, Knowlton et al. 2006, Kwak and Jung 2019, Gardner, Gold et al. 2020). In contrast, deliberative (goal-directed) control results in behavioral flexibility. As such, experimentally, behavior is classified as deliberative or goal-directed if an animal changes its behavior following relevant changes to the action-outcome coupling and/or following a change in the motivational state of the animal which changes the significance of the outcome (Miller, Ludvig et al. 2018).

Prior studies have suggested that animals display prospective behavior with limited training on a task, but under reward conditions that require repeated behaviors, experience leads to a transition toward habitual behavior. In cases where there are un-cued changes to the reward contingencies, both behavioral and neural correlates of decision-making suggest that animals return to deliberative control until the new set of contingencies are learned, and then subsequently re-automate their behavior (Wikenheiser and Redish 2015, Papale, Zielinski et al. 2016, Redish 2016, Hunt, Daw et al. 2021). A well-studied contingency switching task is the left/right/alternation (LRA) task (Gupta, van der Meer et al. 2010, Gupta, van der Meer et al. 2012, Powell and Redish 2014, Regier, Amemiya et al. 2015, Hasz and Redish 2018), which is a form of reversal task. In the LRA task, rodents run through a central track and then turn left or right for food under one of three contingencies. Un-cued changes to the reward contingency (e.g., switching the contingency from left-rewarding to right-rewarding or to requiring alternation to get reward) causes changes to the consequence of taking the left or right action. Behaviorally, procedural control is hypothesized to drive behavioral stereotypy, and prospective control is hypothesized to result in variable behaviors at decision-points (i.e., points at which the animal has to commit to going either left or right). Extensive research on the LRA task suggests that during the initial learning phase, rats elicit variable behaviors at difficult decision-points (such as *vicarious trial and error* (VTE; Tolman (1948), Redish (2016)), orientation and reorientation behaviors) which disappear after extensive experience. After continued experience with a given choice, stereotyped paths emerge — both stereotypy in the path taken to the decision-point, and in the path through the decision-point itself (Schmitzer-Torbert and Redish 2002). After an un-cued change to the reward contingency, behavioral stereotypy disappears and variable behaviors (such as VTE) start to reemerge, which suggests a change from procedural back to prospective behavioral control. As those animals get back into a flow, the behavioral stereotypy gradually reappears, and variable behaviors near disappear. Therefore, with this LRA experimental paradigm, multiple instances of arbitration between prospective and procedural control can be studied (Powell and Redish 2014, Regier, Amemiya et al. 2015, Powell and Redish 2016, Hasz and Redish 2020).

## Algorithmic models of decision-making

*Reinforcement learning* models capture several core features of learning and choice, and thus are widely used in research on decision-making and cognitive processes (Sutton and Barto 2018). In simple tasks, basic reinforcement learning models have been very successful in encapsulating the principles of habitual and prospective (plan-based) control.

### Reinforcement learning algorithms

Algorithmically the distinction between prospective and procedural decision-making is thought to be analogous to that between model-based and model-free reinforcement learning, respectively. *Model-based reinforcement learning* provides a form of planning by maintaining an explicit causal model of the world and using it to choose actions after assessing their likely consequences (Fig. 1 A1. Thus, it enables goal-directed planning. In contrast, *Model-free reinforcement learning* does not maintain an explicit causal model and thus does not allow planning. Instead, actions are assigned values based on their context-dependent history of reward (Fig. 1 B1. The resulting cached policies (stimulus-response habits) are globally adaptive but may not be locally adaptive — for example, trying to go through blocked paths to the goal. Certain implementations of model-free reinforcement learning, such as prediction-error updating and temporal difference learning, have been argued to match well with the midbrain dopamine system responses to expected and unexpected rewards (Schultz, Dayan et al. 1997). While dopamine is also released in tasks believed to use model-based reasoning (Glascher, Daw et al. 2010, Daw, Gershman et al. 2011, McNamara, Tejero-Cantero et al. 2014, Gomperts, Kloosterman et al. 2015), what role dopamine plays in model-based reasoning (planning) remains unknown.

### Hierarchical reinforcement learning models

The effectiveness of reinforcement learning models deteriorates as the size of the learning problem and the complexity of the task grows, thus making basic reinforcement learning solutions incapable of handling rich and naturalistic environments, or problems with large environmental state-spaces and complex action-outcome relationships. One way in which this problem has been addressed is through state abstractions which collapse interrelated environment conditions (e.g. specific locations) into categorized states, treating them as equivalent or interchangeable (Botvinick, Niv et al. 2009, Niv 2009, Botvinick 2012, Diuk, Schapiro et al. 2013). State abstraction decreases the overall number of states the agent must learn about.

These models fall under the construct of *hierarchical reinforcement learning (*Singh 1992*,* Sutton and Barto 2018*)*. Under hierarchical variants, reinforcement learning actions are grouped into 'options', where sequences of actions are bundled together into *action chains* for selection by a controller based on the abstract state (Dezfouli and Balleine 2013, Dezfouli, Lingawi et al. 2014). For example, in the classic hierarchical subgoal schema of traveling to a conference, the first decision is whether to fly or drive. Once one decides to fly, one has to decide which airline to take and when to go. Once one has decided that, one has to decide how to get to the airport, and, similarly, how to get from the airport to one's hotel at the conference. Hierarchical reinforcement learning addresses large scale questions first (fly vs drive) and then breaks that down into smaller scales as necessary (which airline, how to get to the airport).

Importantly, associating action sequences with abstract states means that they can be generalized across contexts (often referred to as '*policy abstraction*'). For example, braiding hair comprises of many individual actions that are then collapsed to a single motor routine that can be repeated for each braid. Using the same example, learning to braid hair can generalize to braiding strings together to make strong ropes. While policy abstraction based decision-making processes are well suited to situations where the reward contingencies and environment statistics remain constant (i.e., cached action chains between subgoals remain optimal), they are poorly suited to circumstances where there are changes to either the reward or state transition probabilities (i.e., cached action chains between subgoals may no longer be optimal, and therefore would necessitate reevaluation). There is growing research across species that suggest that the brain parses ongoing behavior in discrete and bounded segments in line with hierarchical hypothesis of behavioral control (Graybiel 1998, Fujii and Graybiel 2003, Jin and Costa 2010). However, neurobiological instantiations of these strategies are still an active research area.

## Successor representations

Model-based and model-free theories of decision-making represent two extreme ways of calculating the expected reward of taking an action. In addition to model-free and mode-based methods, there are intermediate solutions that rely on learning useful representations. One such approach that leverages aspects of model-based and model-free learning is the *successor representation (*Dayan 1993*,* Gershman 2018*)*. The successor representation can be intuitively thought of as a predictive map that includes not only the current state, but also representations of the likely set of other states that will be visited in the near future (Fig. 1 C1). Therefore, unlike classical methods that iteratively calculate expected reward, successor representations iteratively calculate expected state occupancy. For example, in basic model-free methods the expected reward estimate is updated using the *reward prediction error* (the discrepancy between observed and expected value), while successor representations are updated using an error signal of the discrepancy between observed and expected state occupancy (Gardner, Schoenbaum et al. 2018, Geerts, Gershman et al. 2023). Given its reliance on the statistics of the world, successor representation states that have similar reward estimates can be represented as a single entity during planning. However, it is important to note that successor representations, unlike sequence theories — where a set of states is represented explicitly by the sequence through them — encodes these components in a single component and suggests that reliable parts of the task can be grouped together.

In terms of the efficiency-flexibility trade off, the successor representation lies somewhere in between model-based and model-free algorithms. It has comparable efficiency to model-free reinforcement learning during decision-making because it does not need to take the time to search through possibilities, while maintaining some of the flexibility of model-based algorithms after reward changes. Successor representations can respond quickly when reward contingencies change, but state transition contingencies do not, such as in simple devaluation paradigms, because it separates reward evaluation from representations of state contingency. On the other hand, successor representation responds slowly to changes to the consequences of actions, i.e. when the state transition statistics change, because those consequences are encoded within the successor representation itself and therefore need to be updated.

In multi-step spatial tasks that feature contingency changes to the reward, such as the left/right/alternate task, actions between subgoals would remain stereotyped, as each action continues to reliably lead to the expected state. After reward contingency changes, the successor representation would devalue all the actions taken between reward and the prior successor subgoal, making the sequence less likely to be initiated, but all actions would remain stereotyped when it was (Fig. 1 C2).

## Planning with action-chains

Another option to simplify the planning process is to instead group actions together by chunking them — similar to hierarchical reinforcement learning. Unlike successor representations, where states are grouped together based on their expected occupancy, action chains cache policies between abstract subgoal states (Dezfouli and Balleine 2013, Dezfouli, Lingawi et al. 2014, Cushman and Morris 2015, Holroyd, Ribas-Fernandes et al. 2018) (Fig 1 D1, E1). Compared to non-abstracted model-based planning, action chains that are defined over reliably valued goals makes the problem much more computationally efficient by only requiring search over cached paths to the goal rather than requiring a derivation of the expected reward from search over the full model of the task.

When actions reliably follow each other, and reliably lead to goals (whether they be the end goal, intermediary self-rewarding, or explicitly rewarding subgoals), there are two broad possible models of decision-making: 1) planning over abstract states that are linked with cached chunked actions (Fig. 1 D1) in which the neural representations are about the future, and 2) habit over abstract states that are linked with cached chunked actions (Fig. 1 E1) in which the neural representations are about the current state of the world. In effect, these action-chain models are generalizations of the model-based and model-free decision systems identified above – planning over abstract states is a model-based system with explicit goal representations, but the states are separated by dependable action sequences, while habit is a model-free system where a given situation releases that dependable action-chain. Planning over abstract states contains representations of the goal at the moment of decision (Fig. 1 D3), while habit over abstract state systems will contain representations of the current state of the world (in the task, shown in Figure 1, this will be the side the animal came from, Fig. 1 E3).

An ability to plan between abstract states provides a level of flexibility that the successor representation does not have. Planning between abstract states is most effective when both the value of the cached intra-state action-chains and the transition probabilities between intra-states remain constant. Similar to successor representations, since actions are grouped together, any devaluation can be quickly propagated backwards through the abstract states to the beginning of the trial. This quick propagation allows for decision-making under such a hierarchical representation to be robust against reward devaluations. Behavior with either plan- or habit-based control using abstract states would be very similar (Fig. 1 D2, E2), however we would expect behavioral differences between decision-making using successor representation and decision-making using action-chains and abstract states. With either plan- or habit-based behavioral control, paths through space should be stereotyped, but should change based on changes to the reward contingency. Therefore, each contingency would have its own stereotyped paths. In contrast, with successor representations, the path through the central track should be stereotyped independent of the current contingency, because successor representations are based on the subset of state transitions that are reliable.

### Deep learning

Deep learning has been very successful in using instantaneous representations to learn optimal policy. Unlike classical state-based reinforcements, where the main goal is to learn task representations, deep learning based policy-planning algorithms learn representations that are useful for predicting what the next action should be (Richards, Lillicrap et al. 2019). These methods have been very successful in solving many difficult artificial intelligence problems, such as Atari (Mnih, Kavukcuoglu et al. 2015) and Go (Silver, Schrittwieser et al. 2017). However, because the deep learning process must learn a condensed representation of its own predictive task-space, these implementations require millions of iterations to learn the representations that enable decision-making. The models that show planning-like abilities to navigate complex situations once they have had sufficient training (millions of trials) have been able to abstract super-structured state-spaces that contain second-order information (Niv 2009, Hamrick, Friesen et al. 2020). Once those super-structured state-spaces have been learned, planning-like behaviors can be seen even with purely model-free reinforcement learning algorithms. On the other hand, these processes rely on the stability of the super-structured state-space. Task-representation based planning allows faster learning in novel situations, by generalizing between tasks that can be similarly represented.

## Navigating a complex world

Efficient decomposition of the task into abstract states increases the efficiency and decreases the overall computational burden of deliberative decision-making strategies. Intuitively, these abstract states can be thought of as subgoals. However, finding good and useful subgoals is a difficult problem. In many naturalistic problems the state-space is often high-dimensional making it computationally expensive and time-consuming to explore and evaluate potential subgoals effectively. Moreover, in some environments reward signaling may be sparse, and animals/agents may receive feedback only after reaching the ultimate goal and not for intermediate steps. Without informative rewards along the way, it becomes harder to identify and reinforce useful subgoals that lead to the desired outcome. When the environment is complex and exhibits variations, subgoals that are useful in one part of the state space may not necessarily be useful in other regions. A possible solution to subgoal identification leverages the inherent state connectivity of the task. For example, if the task space is portioned into discrete sets of states that are densely interconnected but sparsely outward connected, it can be useful to identify “hub” states that connect densely connected states (Karuza, Thompson-Schill et al. 2016, Ju and Bassett 2020). The topic of graph segmentation has been largely explored in the context of finding subgoals for hierarchical reinforcement learning, where subgoals correspond to the states that bridge segments of highly interconnected regions of the task (Diuk, Schapiro et al. 2013, Spellman, Rigotti et al. 2015, Schapiro, Turk-Browne et al. 2017).

Subgoal discovery using successor representations use a low dimensional structure of the transition and sampling statistics (Stachenfeld, Botvinick et al. 2017). Intuitively, regions of the task space that is more frequently visited would form the densely interconnected regions mentioned above. Therefore, the new state space could be comprised of these subgoals with cached actions in between. However, this method of subgoal discovery assumes that the precomputed information about the environment statistics and transitions do not change.

Another way to identify subgoals within an environment is to look at the gradient of graph connectivity (Mugan and MacIver 2020, Espinosa, Wink et al. 2022). Here, subgoals correspond to locations of high gradient of environmental connectivity. Unlike successor representations that rely on visitation statistics, this graph-gradient approach uses the actual structure of the environment and is therefore agnostic to how those transition statistics might be obtained (i.e., what actions an agent initially takes to create an occupancy map). These transition regions act as abstract states over which planning occurs, and between these transition regions, actions are selected based on either previously learned action sequences or shortest path — a strategy commonly observed in animals. Because subgoals in this model are selected based on structure of the environment, this approach yields favorable outcomes even in highly volatile environments where the reward structure is not static (Shamash, Olesen et al. 2021, Espinosa, Wink et al. 2022, Krausz, Comrie et al. 2023, Shamash, Lee et al. 2023).

It is important to note that whatever approach is taken to create an abstract state-space, the geometry of the space influences the value of employing either planning or habit strategies. This suggests that environments that feature many distributed decision points will encourage decision-making that relies more on planning than habit, and that such tasks would develop habitual responses slower. Given that deliberating over choices is computationally expensive, these environments would also result in higher memory demands.

## Neural correlates of decision-making

### Hippocampus

Research over the past decade suggests that hippocampus is a crucial component of the network supporting memory function and navigation (O'Keefe and Nadel 1978, Eichenbaum 1993, Redish 1999, Preston and Eichenbaum 2013, Lisman 2015). Hippocampal manipulations and lesions in rodents have shown impairments in the ability to recognize familiar environments (O'Keefe, Nadel et al. 1975), to navigate to hidden spatial goals (Morris, Garrud et al. 1982), and overall deficits in working memory (Rawlins 1985, Fortin, Agster et al. 2002). Similar complementary results have been found in humans, such as the deficits in spatial memory that result in impairments to the ability to flexibly navigate through the world (Maguire, Nannery et al. 2006). Electrophysiological investigations of the hippocampus and associated regions in rodents have identified ‘place cells’ (O'Keefe and Dostrovsky 1971), and other neural elements that support memory and spatial cognition. Place cells, which are commonly associated with the pyramidal neurons in the CA1 and CA3 layers of the hippocampus are named as such because they exhibit stable and spatially constrained firing fields (i.e., place fields) (O'Keefe and Nadel 1978, Ekstrom, Kahana et al. 2003, Ulanovsky and Moss 2007). Notably, place cells have been shown to encode environmental parameters such as environment geometry (Lever, Wills et al. 2002, Leutgeb, Leutgeb et al. 2005), color (Jeffery and Anderson 2003), size (Fenton, Kao et al. 2008, Zhang, Rich et al. 2023), and the arrangement of connected compartments (Duvelle, Grieves et al. 2021), and place maps have been shown to represent new goal locations (Knierim, Kudrimoti et al. 1995, Lee, LeDuke et al. 2018). All of these properties further buttress the hippocampal role in environmental model building.

Given that place cells encode information about the environment, goal(s), and the trajectory of the animal through it, spike sequences can be extracted to decode what the spike trains are encoding about the animal's current state (Zhang, Ginzburg et al. 1998, Johnson and Redish 2007, Deng, Liu et al. 2016, Denovellis, Gillespie et al. 2021). Probabilistic decoding, sequence analysis, and other methods applied to direct measurements of hippocampal place cells have been used to elucidate non-local representations that suggest imagination of future outcomes, as would be predicted from a hippocampal role in planning (Fig. 1 A3) (Johnson and Redish 2007, Kay, Chung et al. 2020).

Recently, it has been proposed that hippocampal place cells may be encoding a successor representation rather than a cognitive map of the environment (Fig. 1 C3) (Stachenfeld, Botvinick et al. 2017). Studies have shown that changes to the transition probabilities from state to state, such as insertion of barriers, has been shown to create changes in place cell firing rates and place fields, which suggests local remapping due to topological changes (Chen, Gomperts et al. 2014, Dabaghian, Brandt et al. 2014, Wu and Foster 2014, Duvelle, Grieves et al. 2021). Under the successor representation hypothesis, such local remapping would be related to the changes in the environmental connectivity (both due to changes to the low dimensional representation of the environment, and to the expected occupancy of each state). However, the way in which successor representations condense the state-space seems to be in conflict with the decoded state observed during hippocampal non-local activity, where each non-local sequence serially encodes states to assess future outcomes, e.g. CA1 sweeps at the choice point (Fig. 1 A3). The existence of this serial assessment by hippocampus suggests model-based strategies rather than successor representations during prospective decision-making.

### Prefrontal cortex

While studies have found that the hippocampus is at the center of state-space representations, prefrontal cortex is hypothesized to encode the 'problem task space' (Hok, Save et al. 2005, Durstewitz, Vittoz et al. 2010, Powell and Redish 2014), goals (Hok, Save et al. 2005), subgoals (Botvinick 2008), and strategy (Powell and Redish 2016, Hasz and Redish 2020).

Within the prefrontal cortex, anatomical findings have revealed subdivisions along the dorsal-ventral axis that are thought to regulate different modes of decision-making (Kesner and Churchwell 2011, Laubach, Amarante et al. 2018, Diehl and Redish 2023). In the rodent, prefrontal cortex ranges from anterior cingulate cortex (ACC) at its most dorsal, to the prelimbic cortex (PL) — where new work suggests that there may be a functional subdivision that exists along the dorsal-ventral axis of PL (Diehl and Redish 2023) — to the infralimbic cortex (IL) at the most ventral (Heidbreder and Groenewegen 2003, Uylings, Groenewegen et al. 2003, Hoover and Vertes 2007).

Dorsal prefrontal manipulation studies have shown that disruptions to the prelimbic cortex in rats disrupts rule-based decision-making and synchrony/transfer of information between hippocampus and prefrontal cortex (Shaw, Watson et al. 2013, Padilla-Coreano, Canetta et al. 2019, Schmidt, Duin et al. 2019, Broschard, Kim et al. 2021, Kidder, Miles et al. 2021). The excitation of dorsal prefrontal networks has been shown to increase high-frequency gamma power, which often is interpreted as increased synchronization between prefrontal cortex and connected subcortical regions (orbitofrontal cortex, ventral striatum) that are implicated in value-based prospective decision-making (van der Meer,

Kalenscher et al. 2010, Siegel, Donner et al. 2012, Spellman, Rigotti et al. 2015, Ferenczi, Zalocusky et al. 2016).

Optogenetic studies have shown that inhibition of IL either prevented the formation of habits or blocked recently acquired habits (Smith, Virkud et al. 2012, Smith and Graybiel 2013). While IL lacks direct connections with regions that are hypothesized to be involved in habit expression (Hurley, Herbert et al. 1991), it needs to be intact for those habits to be expressed (Coutureau and Killcross 2003, Killcross and Coutureau 2003). Why this is true remains unknown, but it has been suggested that IL may exert executive-level online control in the selection of habits through its prefrontal-limbic system connections, while the representation of the selected habit itself resides in other sensorimotor networks (Killcross and Coutureau 2003, Hitchcott, Quinn et al. 2007, Smith and Graybiel 2013, McLaughlin, Diehl et al. 2021).

Taken together, our current understanding about prefrontal cortex's role in decision-making suggests a competition between dorsal and ventral prefrontal cortex, with dorsal prefrontal cortex more involved in deliberative systems, and ventral prefrontal cortex more involved in procedural systems (Killcross and Coutureau 2003, Daw, Niv et al. 2005, Vidal-Gonzalez, Vidal-Gonzalez et al. 2006, Smith and Graybiel 2013, Mukherjee and Caroni 2018, Diehl and Redish 2023).

### Dorsolateral, dorsomedial, and ventral striatum

Although there is no real clear boundary across the dorsal-ventral or medial-lateral axes of the striatum, striatal afferent and efferent projections (both corticostriatal and thalamostriatal projections) reveal clear gradients separating striatum into subdivisions along those dorsal-ventral and medial-lateral (as well as anterior-posterior) axes (Gerfen 1984, Brown, Bullock et al. 1999, Heilbronner, Rodriguez-Romaguera et al. 2016, Hunnicutt, Jongbloets et al. 2016, Peters, Fabre et al. 2021). Current theories suggest the striatum has three functional subdivisions: a sensorimotor division that corresponds to dorsolateral striatum, a cognitive processing division (e.g., place strategy, working memory) that corresponds to dorsomedial striatum, and a reward processing limbic division that corresponds to ventral striatum (Joel, Niv et al. 2002, Knutson, Taylor et al. 2005, Balleine, Liljeholm et al. 2009, van der Meer and Redish 2009, Thorn, Atallah et al. 2010, van der Meer, Johnson et al. 2010, van der Meer, Kalenscher et al. 2010, Regier, Amemiya et al. 2015, Hunnicutt, Jongbloets et al. 2016, Gahnstrom and Spiers 2020).

The dorsolateral striatum receives sensorimotor-related information and is thought to store action plans for habit learning as a result of its anatomical connections (Yin and Knowlton 2006). There are numerous studies that have shown that inhibitory manipulations to the dorsolateral striatum disrupts stereotyped habit behaviors (Yin, Knowlton et al. 2004, Yin, Knowlton et al. 2006). Recordings from the dorsolateral striatum have shown punctate bursting activity that is hypothesized to encode the beginning and end of stereotyped action sequences (Fig. 1 B3 left panel) (Graybiel 1998, Jog, Kubota et al. 1999, Barnes, Kubota et al. 2005, Graybiel 2008, Jin and Costa 2010, Jin, Tecuapetla et al. 2014). (However, it is still an open question as to how intermediary decision points may change this firing pattern, and whether action sequences in between these intermediary points that occur between trial start and end are also represented by punctate bursting activity, e.g. increased striatal activity at the turns T1-T4 in a multiple T-maze (Fig. 1 B3 right panel)).These bursts are hypothesized to indicate that decisions are being made over cached action-chains under procedural control. Recent studies have shown that these task-initiation bursts occur shortly after the animal initiates its movement. Decoding of striatal activity during these bursts suggests that they reflect the current state and recent past experiences more than future options or planning processes (Cunningham, Regier et al. 2021).

The more ventromedial part of striatum receives associative and motivational information, such as information from hippocampus, orbitofrontal cortex, and prelimbic cortex, all implicated in deliberative control (Voorn, Vanderschuren et al. 2004, van der Meer, Johnson et al. 2010, van der Meer and Redish 2010). Lesions of the dorsomedial striatum has been shown to disrupt aspects of spatial learning, and a preference for cue-response habit strategy (Pooters, Gantois et al. 2016). In contrast, lesions to the dorsolateral striatum does not affect learning, but rather decreases the use of procedural strategies (Yin, Knowlton et al. 2004). Ventral striatum has been shown to contain anticipatory reward-related firing (Lavoie and Mizumori 1994, Schultz, Dayan et al. 1997). Importantly, ventral striatal neurons have been shown to fire selectively in anticipation or during reward delivery, making ventral striatal neural coding plausibly about the relevant elements of goal-directed tasks along with its motivational component (Khamassi, Mulder et al. 2008, van der Meer and Redish 2010, van der Meer and Redish 2011, van der Meer and Redish 2011).

Striatum's functional subdivisions — demarcated based on its heterogeneous projection patterns to the pallidum, substantia nigra, and the thalamus to the cortical areas — form parallel but distinct loops (Groenewegen, Galis-de Graaf et al. 1999, Thorn, Atallah et al. 2010). Ito and Doya (2011) suggest that the gradient in the input-output projections of the striatum may be for implementing hierarchical reinforcement learning, where ventral subdivisions valuate top-level goals, and dorsal subdivisions select and perform action chains. Under this theory the prefrontal cortex would still be viewed as a meta controller that provides top-level state information and sets strategies based on the received information (Hikosaka, Nakahara et al. 1999, Botvinick 2008).

## Competition between deliberative and habitual decision-making in navigationally complex environments

A broad range of research into neural and behavioral correlates of decision-making suggests that the brain contains multiple systems for generating adaptive strategies such as habitual and deliberative control (Daw, Niv et al. 2005). However, the existence of multiple controllers raises questions: On what does the arbitration between the two systems depend? What role does the task and environment structure play in advantaging the two systems?

Mathematically, any problem in which the task/state-space and the transitions between them are known can be transformed into a generalized relational structure, i.e., a 'cognitive map' (Tolman 1948, O'Keefe and Nadel 1978) Neural instantiations of the cognitive map have been extensively studied in spatial domains, where map-like representations have been found in specific neurons in the hippocampal-entorhinal system that have activity patterns that seem to encode different aspects of the task structure, e.g. place cells (O'Keefe and Dostrovsky 1971, Solstad, Moser et al. 2006). There is increasing evidence of non-spatial relational organization of knowledge, which extends the spatial cognitive map to a general 'concept' space (Aronov, Nevers et al. 2017, Behrens, Muller et al. 2018,

Whittington, Muller et al. 2019, Baram, Muller et al. 2020). Together, our current understanding suggests that there is some representation of the structure of the problem.

Prior research in many different disciplines that deal with relational structures (e.g., chemistry, discrete mathematics) have used graph entropy as a way to determine and quantify the structural complexity and information content of graphs (Trucco 1956, Newman, Barabási et al. 2006, Dehmer and Mowshowitz 2011). To determine the structural or topological complexity of an environment, we first discretized the inner navigation space of the maze into *n* spatial nodes that were connected by a single action (Mugan and MacIver 2020, Mugan, Hoffman et al. 2022). Starting from each node we created *n* generalized trees, with each root node being one of the nodes in the discretized maze. The structural complexity of a tree is defined as the Shannon entropy of the distribution of the number of nodes at each level of the tree (Bonchev 2009). The total structural complexity of the set of generalized trees is defined as the total information content across all the equivalent generalized trees that can be created from a given navigational environment. Intuitively this means that a graph has maximal entropy if it possesses the same number of nodes at each level — i.e., each state can only be reached from one state, and therefore the space is harder to navigate, and mistakes are more costly.

Prior studies have shown that in tasks that feature un-cued changes to reward contingencies, both deliberative and procedural behaviors and their neural correlates can be observed. As described previously, changes to reward contingencies have been proposed to lead to procedural responses becoming deliberative until the new set of contingencies are learned, after which behavior is again hypothesized to be driven by the procedural system, leading to re-automation. Prior work has shown that environmental topology and complexity should differentially advantage different decision-making systems, with more complex environments favoring deliberation (Mugan and MacIver 2020, Mugan, Hoffman et al. 2022). To study how environmental topology and complexity interacts with these different decision-making systems, we re-analyzed data from a task with a changing central track. We measured how central track topology impacts decision-making, and the arbitration between deliberative and procedural decision-making systems. We reanalyzed behavior and recordings from hippocampus and dorsolateral striatum during a contingency-switching task with a variable central path, in the light of the novel graph-theoretic measure of maze complexity.

In the left/right/alternate (LRA) task, rats run through a central track and then turn left or right for food under one of three contingencies. Under the left contingency, food reward is presented on the left side, under the right contingency the food reward is presented on the right side, and under the alternation contingency, the food reward is presented on the opposite side of the previous run — i.e., the rats must alternate between left and right to receive food reward (Fig. 2A). To study multiple instances of changes to behavioral strategy, each session included a change in reward contingency roughly halfway through the session that was set in the absence of external cues. Re-analysis was conducted on neural ensembles ($n_{\text{Sessions}}$ = 50) that were recorded from dorsolateral striatum (6 rats) or hippocampus (5 rats) using tetrodes (Regier, Amemiya et al. 2015, Amemiya and Redish 2016, Amemiya and Redish 2018).

## Deliberative decision-making in environments with varying central path complexity

Deliberative decision making is hypothesized to result in variable behaviors (Redish 2016). In spatial tasks behavioral variability can be observed throughout the entirety of the task, including variability in the paths taken as the rats run through the central track. We calculated the variance in the paths taken through the central track and found that paths were more variable on more complex tracks (Fig. 2B, C), which suggests that the complexity of the central track may modulate deliberative behaviors.

A behavioral correlate of deliberative decision making in spatial tasks is variable behavior at decision-points, such as *vicarious trial and error* (VTE) behaviors, often seen at the choice point (i.e., T junction in the maze where the rats have to make a high-cost decision to either go left or right for a food reward). During VTE, rats pause at difficult decision-points and orient toward a goal (e.g. look/move towards the left or right arm), and then reorient back and forth (Fig. 2D top) (Tolman 1939, Tolman 1948, Johnson and Redish 2007, Redish 2016, George, Stout et al. 2023). As such, non-VTE behaviors feature a smooth path through the choice point with no pauses, while VTE behaviors feature pauses and changes in the angular velocity of the path (Fig. 2D bottom). Complex environments, which feature higher variance in the paths taken through the central track (Fig. 2C), also showed an increase in the proportion of VTE events per session (Fig. 2E).

A neurophysiological correlate of deliberative decision making is cell assemblies in the hippocampus sweeping forward ahead of the animal, serially assessing the outcomes of possible choices (Johnson and Redish 2007, Kay, Chung et al. 2020). Experiments have found that during times when animals would be expected to be using prospective (deliberative) decision-systems, representations sweep forward during each hippocampal theta cycle (Fig. 1 A3). Sweep length has been found to go longer when goals are farther away, as well as going longer during VTE events (Wikenheiser and Redish 2015, Wikenheiser and Redish 2015).

Current theories suggest that theta cycles consist of two components: 1) current location representation during the first-half of theta, and 2) nonlocal forward information representation during the second-half of theta (Fig. 1 A3) (Lisman and Redish 2009, Schmidt, Duin et al. 2019). Intuitively this can be thought of as sweeps starting from local position and moving ahead of the animal during the second half of theta. The asymmetry in the theta cycle is a result of the differences in the duration between first and second halves. To assess the changes to local and nonlocal representations of space with respect to environmental complexity we calculated the asymmetry index of individual theta cycles. The asymmetry index was quantified as the log ratio of the duration of the second half and first half of theta. Fascinatingly, the first of the theta cycle does not change much from cycle to cycle, and changes in the asymmetry tend to be driven by changes in the length of the second half (Amemiya and Redish 2018, Schmidt, Duin et al. 2019). A theta cycle that has a more negative asymmetry index has a shorter second-half, and therefore is expected to represent more local information. Conversely, a theta cycle with a more positive asymmetry index has a longer second-half, and therefore is expected to represent more nonlocal information. These sweeps are moving farther ahead of the animal during the second half of theta. In line with previous research, the asymmetry index is positively skewed suggesting more asymmetric cycles, which suggests more local firing (Fig. 2H) (Belluscio, Mizuseki et al. 2012, Amemiya and Redish 2018, Schmidt, Duin et al. 2019). Notably, we found that changes to the central path topology changed theta symmetry. We see that higher central path variability results in a shift towards more positive asymmetry in the theta cycles (Fig. 2H), which have second-halves with longer duration (Fig. 2I), indicating an increase in the non-local activity.

Another potential neural substrate of deliberate decision-making is hippocampal replay during sharp-wave ripple (SWR) events, in which certain hippocampal cells are sequentially activated in millisecond timescales (Diba and Buzsaki 2007, Buzsaki 2015, Foster 2017). These sequences represent recent experiences (Gillespie, Astudillo Maya et al. 2021), novel (Gupta, van der Meer et al. 2010), and future paths (Olafsdottir, Barry et al. 2015). Interruption of sharp-wave ripples leads to deficits in learning and performance (Jadhav, Kemere et al. 2012). Recent work has shown that the fraction of long-duration sharp-wave ripples during awake immobility increases with task memory demands and is important for memory consolidation (Fernandez-Ruiz, Oliva et al. 2019). In line with the idea that complex environments require a higher memory load, we also observe significantly longer SWRs in complex environments during awake immobility (Fig. 2J). While not significant, there was a trend that suggests there are more elongated SWR activity coded as chains of ripples (ripple bursts (e.g., doublets and triplets of SWRs)) in more complex environments (500ms window) (Fig. 2K). This is in line with previous research that shows that number of ripple bursts increase with increased memory load (Fernandez-Ruiz, Oliva et al. 2019), and in tasks that require extended place cell sequences (e.g., due to a longer track) (Davidson, Kloosterman et al. 2009, Yamamoto and Tonegawa 2017, Judak, Chiovini et al. 2022). These results together suggest that complex environments require increased memory load which are reflected in increased SWR activity, but whether that SWR activity plays a prospective role in planning or merely a consolidation role remains unknown. For example, causal manipulations of SWR have found consistent disruption in memory consolidation processes (Ego-Stengel and Wilson 2010, Gridchyn, Schoenenberger et al. 2020).

## Procedural decision-making in environments with varying central path complexity

While deliberative processes are often associated with variable behaviors, procedural processes are often associated with highly stereotyped behaviors. Behavioral stereotypy through the central path was defined as the reciprocal of the Euclidean distance between each lap to every other lap (Fig. 2F). We find that, in both simple and complex environments, behavior was stereotyped after subjects had been running a contingency for a number of laps, prior to the contingency switch, as would be expected from behavior under the procedural system (Fig. 2G). Contingency switches were followed by a decrease in behavioral stereotypy, as would be expected from behavior under the deliberative system. However, while there was a similar progression of stereotypy aligned to reward contingency switches in both simple and complex environments, overall stereotypy was significantly lower in complex environments, which suggests slower automation (Fig. 2G).

Dorsolateral striatum is hypothesized to be necessary for habitual performance, and striatal ensemble activity shows changes to neuronal activity with experience on a task (Graybiel 2008, Smith and Graybiel 2016). In spatial tasks, with limited experience, there is prominent striatal activity throughout the entirety of each lap. However, with increased experience, striatal activity becomes increasingly prominent at the beginning and end of each lap, with a notable decrease in activity through the middle of the journey (Barnes, Kubota et al. 2005, Jin and Costa 2010, Jin, Tecuapetla et al. 2014). This phenomenon is usually referred to as *task bracketing* and is thought to be indictive of decisions being made over chunked actions.

Consistent with previous studies, we quantified task bracketing as a *task-bracketing index*, measured as the normalized ratio between differences in mean striatal firing rates at the start and at the end of the maze over the standard deviation of the mean firing rates everywhere else. A high task-bracketing index is interpreted to reflect striatal ensembles developing a hierarchical representation of action boundaries. Overall, we found that task-bracketing index developed over the course of a stable contingency, but the development of task-bracketing was slower in complex environments during the first contingency (Fig. 2L; top panel). There was no significant effect of environmental complexity on the task-bracketing index in the second contingency (Fig. 2L; bottom panel). This suggests that the initial habituation is slower in complex environments, and that complex environments may require more experience to learn the task.

Clear behavioral and neurophysiological markers of deliberative and procedural decision-making, as well as the arbitration between the two, exist in our left/right/alternate contingency switching task. Using a novel graph-theoretic measure we quantified the complexity of the navigational sequence to study how environmental topology may be affecting these different decision-making systems. Our reanalysis of previously published results (Regier, Amemiya et al. 2015, Amemiya and Redish 2016, Amemiya and Redish 2018) shows that environmental complexity impacts decision-making processes. Deliberation may be particularly crucial in complex environments. Consistent with this we see slower automation in complex environments. Taken together these data suggest that complex environments may be requiring a higher cognitive memory load and a more detailed cognitive map.

## Conclusion

Decision-making and learning are central for the animals to be able to adapt to the external environment and changes to it. Current theories suggest that there are multiple different decision systems that are in competition with each other. The classic dichotomy has been between prospective (deliberative/planning) and procedural (habit) behavioral control systems (Daw, Niv et al. 2005, Hunt, Daw et al. 2021). Both decision-making systems have been successfully modeled by reinforcement learning methods. Commonly, the behavior under the prospective system has been algorithmicized by model-based theories (Fig. 1 A1), and the behavior under the procedural system has been algorithmicized by model-free theories (Fig. 1 B1) (Niv 2009, Sutton and Barto 2018). While these theories have been very successful in explaining neural and behavioral data, several new theories have been proposed that meld aspects of the classic prospective and procedural theories: successor representations (Fig. 1 C1) (Dayan 1993), planning over chunked action chains (Fig. 1 D1) , and caching chunked action-chains (Fig. 1 E1) (Dezfouli and Balleine 2013, Dezfouli, Lingawi et al. 2014). Successor representations store repeated world paths probabilistically so as to simplify the planning across them. The chunked action-chain theories propose that the actions are chunked such that decisions are made over sequences of actions. Importantly, successor representations do not include representations of the value of the outcome, enabling one to make fast changes when there are changes to the reward contingency. In contrast, chunked action-chain theories represent the entire path to the next outcome which can then either be evaluated by imagination of outcomes or through cached valuation. The different predictions each of these algorithms make can be experimentally tested and would predict different behavioral and neural manifestations. Contingency switching tasks with varying environmental structures, such as the LRA variants (e.g., multiple T-maze, Hebb Williams maze (Fig. 2B)), are particularly well-suited to differentiate between these different decision-making algorithms.

The different decision-making algorithms predict different behavioral outcomes based on the internal structure of the maze. Model-based algorithms (planning) predict that navigation through the central track would be variable with likely vicarious trial and error events at the high-cost decision point (choice point) (Fig. 1 A2). Model-free algorithms (habit) predict that with experience variable paths should become stereotyped (Fig. 1 B2). Successor representation theory suggests that repeated paths can become represented as a single entity so that planning can occur over them. Behaviorally, successor representation predicts that any stereotyped path that develops should be maintained across reward contingency changes (Fig. 1 C2). Our behavioral results show that stereotypy is maintained across the switches (i.e., the left and right contingency have the same stereotyped path through the central track); however, this stereotypy is not maintained across the switch in reward contingency (Fig. 2G), which suggests that successor representation is not the primary type of behavioral control that occurs in this task. Neurophysiologically, current successor representation theories suggest that dorsal hippocampus place cells should encode a low-dimensional representation of space (Stachenfeld, Botvinick et al. 2017, Geerts, Gershman et al. 2023). This suggests interior complexity should change place field characteristics, such as skewness, size, and lead to overrepresentation/clustering of place fields near doorways or tight turns (Fig. 1 C3).

Making decisions over chunked actions requires a hierarchical representation of the space. The changes to the structure of the interior navigational sequence may create differently-chunked representations, in which complex environments favor finer state representations. Planning over these chunked actions would therefore predict variable behaviors across these chunks, but highly stereotyped behaviors within each chunk (Fig. 1 D1). For example, there may be seemingly superfluous exploration of different states, but actions through those abstract states would be stereotyped. In contrast, habit over chunked actions would predict stereotypy through the entire path. However, unlike successor representation, this theory predicts that stereotyped paths should change between contingencies (Fig. 1 D1, E1).

The different decision-making algorithms also make different predictions about neurophysiology, particularly as they relate to the complex central path. Non-local hippocampal activity, in which place cells alternately sweep to the next goal with each theta process, has been often theorized to reflect the planning process (Fig. 1 A3) (Redish 2016). Historically vicarious trial and error behaviors or hippocampal sweeps have not been observed at low-cost subgoals, however, they have not been studied in environments with complex interior navigation sequences with potentially changing subgoal locations.

Model-free algorithms are often associated with changes to the dorsolateral striatum ensemble activity. Commonly, striatal bursting activity at the start of a bout is hypothesized to indicate the start of cached action sequences (Fig. 1 B3) (Jog, Kubota et al. 1999). However, it is still an open question as to how striatal activity may change when there are intermediate subgoals, and how those subgoals may be represented. For example, are action sequences separately cached between subgoals? Are the start and end of cached action sequences between subgoals also represented by striatal bursting activity (similar to the original idea of task-bracketing but occurring at intermediate stages of the task)? Prior work on a multiple T-maze which featured multiple turns before the choice point found an increase in striatal activity at the maze turns (Fig. 1 B3 right column; indicated by T1-T4), which suggests intermediate representations of start and end of action bouts by striatal cells (van der Meer, Johnson et al. 2010). Systematic changes and manipulation of the interior navigation sequence, in between start and reward sites (i.e., creation of intermediate unrewarded subgoals), would provide insight into how actions may be chunked in between decision-points, and how dorsolateral striatal activity may change with differently distributed decision-points (e.g., Fig. 1 B3 right).

Prior work has shown that dorsolateral striatum cells show task-initiation bursts before stereotyped paths. These bursts are usually taken as indicative of decisions being made over chunked actions (Fig. 1 D1, E1). Cunningham, Regier et al. (2021) applied novel decoding measures to examine what was represented during these bursts. The planning over chunked actions theory (Fig. 1 D1) predicts that activity should reflect and be predictive of the future (a prospective bias), while the habit over cached actions theory (Fig. 1 E1) predicts that activity should reflect the current state and recent past experiences (a retrospective bias). Consistent with previous work, Cunningham, Regier et al. (2021) found that these task-initiation bursts occurred shortly after the animal initiated its movement and coincided with 50 Hz (gamma) striatal oscillations that increased as the subsequent paths became more stereotyped (hypothesized to reflect the development of habit, and the transfer of behavioral control from deliberative to habitual systems). Interestingly, however, they found that these bursts showed a retrospective bias, supporting habit over cached action sequences in lieu of planning theories. While there was a significant retrospective bias in striatal representations, they found a very broad distribution that stretched into the prospective space. A task with better-structured subgoals and better-controlled environmental complexity and topology is needed to more definitively separate the two theories.

Therefore, both behavioral and neurophysiological predictions that each decision-making system makes, and how different decision-making systems interact may be better examined when we consider more naturalistic and complex environments. These theories all point to different neurophysiological and behavioral predictions, particularly as it relates to changing navigational complexity. Studying the impact of environmental structure on different decision-making systems may provide novel insight into the internal dynamics of interacting but distinct brain regions that produce adaptive behaviors.

## Contributions

**Ugurcan Mugan:** Conceptualization, Formal Analysis, Writing – original draft, Writing – review and editing.

**Seiichiro Amemiya:** Hippocampus data collection.

**Paul S. Regier:** Dorsolateral striatum data collection.

**A. David Redish:** Conceptualization, Supervision, Funding acquisition, Investigation, Writing – original draft, Project Administration, Writing – review and editing.

## Acknowledgments

We thank K Seeland, C Boldt, A Sheehan for technical assistance as well as members of the Redish lab for useful discussion.

## References

Amemiya, S. and A. D. Redish (2016). "Manipulating Decisiveness in Decision Making: Effects of Clonidine on Hippocampal Search Strategies." J Neurosci **36**(3): 814-827.

Amemiya, S. and A. D. Redish (2018). "Hippocampal Theta-Gamma Coupling Reflects State-Dependent Information Processing in Decision Making." Cell Rep **22**(12): 3328-3338.

Aronov, D., R. Nevers and D. W. Tank (2017). "Mapping of a non-spatial dimension by the hippocampal-entorhinal circuit." Nature **543**(7647): 719-722.

Averbeck, B. and J. P. O'Doherty (2022). "Reinforcement-learning in fronto-striatal circuits." Neuropsychopharmacology **47**(1): 147-162.

Balleine, B., C. Garner and A. Dickinson (1995). "Instrumental outcome devaluation is attenuated by the anti-emetic ondansetron." The Quarterly Journal of Experimental Psychology Section B **48**(3b): 235-251.

Balleine, B. W. and A. Dickinson (1998). "Goal-directed instrumental action: contingency and incentive learning and their cortical substrates." Neuropharmacology **37**(4-5): 407-419.

Balleine, B. W. and A. Dickinson (2005). "Effects of outcome devaluation on the performance of a heterogeneous instrumental chain." International Journal of Comparative Psychology **18**(4).

Balleine, B. W., M. Liljeholm and S. B. Ostlund (2009). "The integrative function of the basal ganglia in instrumental conditioning." Behav Brain Res **199**(1): 43-52.

Baram, A. B., T. H. Muller, H. Nili, M. Garvert and T. E. J. Behrens (2020). "Entorhinal and ventromedial prefrontal cortices abstract and generalise the structure of reinforcement learning problems." bioRxiv.

Barnes, T. D., Y. Kubota, D. Hu, D. Z. Jin and A. M. Graybiel (2005). "Activity of striatal neurons reflects dynamic encoding and recoding of procedural memories." Nature **437**(7062): 1158-1161.

Behrens, T. E. J., T. H. Muller, J. C. R. Whittington, S. Mark, A. B. Baram, K. L. Stachenfeld and Z. Kurth-Nelson (2018). "What Is a Cognitive Map? Organizing Knowledge for Flexible Behavior." Neuron **100**(2): 490-509.

Belluscio, M. A., K. Mizuseki, R. Schmidt, R. Kempter and G. Buzsaki (2012). "Cross-frequency phase-phase coupling between theta and gamma oscillations in the hippocampus." J Neurosci **32**(2): 423-435.

Bonchev, D. G. (2009). "Information theoretic complexity measures." Encyclopedia of Complexity and Systems Science **5**: 4820-4838.

Botvinick, M. M. (2008). "Hierarchical models of behavior and prefrontal function." Trends Cogn Sci **12**(5): 201-208.

Botvinick, M. M. (2012). "Hierarchical reinforcement learning and decision making." Curr Opin Neurobiol **22**(6): 956-962.

Botvinick, M. M., Y. Niv and A. G. Barto (2009). "Hierarchically organized behavior and its neural foundations: a reinforcement learning perspective." Cognition **113**(3): 262-280.

Breland, K. and M. Breland (1961). "The misbehavior of organisms." American psychologist **16**(11): 681.

Broschard, M. B., J. Kim, B. C. Love, E. A. Wasserman and J. H. Freeman (2021). "Prelimbic cortex maintains attention to category-relevant information and flexibly updates category representations." Neurobiol Learn Mem **185**: 107524.

Brown, J., D. Bullock and S. Grossberg (1999). "How the basal ganglia use parallel excitatory and inhibitory learning pathways to selectively respond to unexpected rewarding cues." J Neurosci **19**(23): 10502-10511.

Buckner, R. L. and D. C. Carroll (2007). "Self-projection and the brain." Trends Cogn Sci **11**(2): 49-57.

Buzsaki, G. (2015). "Hippocampal sharp wave-ripple: A cognitive biomarker for episodic memory and planning." Hippocampus **25**(10): 1073-1188.

Chen, C. S., E. Knep, A. Han, R. B. Ebitz and N. M. Grissom (2021). "Sex differences in learning from exploration." Elife **10**: e69748.

Chen, Z., S. N. Gomperts, J. Yamamoto and M. A. Wilson (2014). "Neural representation of spatial topology in the rodent hippocampus." Neural Comput **26**(1): 1-39.

Coutureau, E. and S. Killcross (2003). "Inactivation of the infralimbic prefrontal cortex reinstates goal-directed responding in overtrained rats." Behav Brain Res **146**(1-2): 167-174.

Cunningham, P. J., P. S. Regier and A. D. Redish (2021). "Dorsolateral Striatal Task-initiation Bursts Represent Past Experiences More than Future Action Plans." J Neurosci **41**(38): 8051-8064.

Cushman, F. and A. Morris (2015). "Habitual control of goal selection in humans." Proc Natl Acad Sci U S A **112**(45): 13817-13822.

Dabaghian, Y., V. L. Brandt and L. M. Frank (2014). "Reconceiving the hippocampal map as a topological template." Elife **3**: e03476.

Davidson, T. J., F. Kloosterman and M. A. Wilson (2009). "Hippocampal replay of extended experience." Neuron **63**(4): 497-507.

Daw, N. D., S. J. Gershman, B. Seymour, P. Dayan and R. J. Dolan (2011). "Model-based influences on humans' choices and striatal prediction errors." Neuron **69**(6): 1204-1215.

Daw, N. D., Y. Niv and P. Dayan (2005). "Uncertainty-based competition between prefrontal and dorsolateral striatal systems for behavioral control." Nat Neurosci **8**(12): 1704-1711.

Daw, N. D., J. P. O'Doherty, P. Dayan, B. Seymour and R. J. Dolan (2006). "Cortical substrates for exploratory decisions in humans." Nature **441**(7095): 876-879.

Dayan, P. (1993). "Improving Generalization for Temporal Difference Learning: The Successor Representation." Neural Computation **5**(4): 613-624.

Dayan, P., Y. Niv, B. Seymour and N. D. Daw (2006). "The misbehavior of value and the discipline of the will." Neural networks **19**(8): 1153-1160.

Dehmer, M. and A. Mowshowitz (2011). "A history of graph entropy measures." Information Sciences **181**(1): 57-78.

Deng, X., D. F. Liu, M. P. Karlsson, L. M. Frank and U. T. Eden (2016). "Rapid classification of hippocampal replay content for real-time applications." J Neurophysiol **116**(5): 2221-2235.

Denovellis, E. L., A. K. Gillespie, M. E. Coulter, M. Sosa, J. E. Chung, U. T. Eden and L. M. Frank (2021). "Hippocampal replay of experience at real-world speeds." Elife **10**.

Dezfouli, A. and B. W. Balleine (2013). "Actions, action sequences and habits: evidence that goal-directed and habitual action control are hierarchically organized." PLoS Comput Biol **9**(12): e1003364.

Dezfouli, A., N. W. Lingawi and B. W. Balleine (2014). "Habits as action sequences: hierarchical action control and changes in outcome value." Philos Trans R Soc Lond B Biol Sci **369**(1655).

Diba, K. and G. Buzsaki (2007). "Forward and reverse hippocampal place-cell sequences during ripples." Nat Neurosci **10**(10): 1241-1242.

Diehl, G. W. and A. D. Redish (2023). "Differential processing of decision information in subregions of rodent medial prefrontal cortex." Elife **12**.

Diuk, C., A. Schapiro, N. Córdova, J. Ribas-Fernandes, Y. Niv and M. Botvinick (2013). "Divide and conquer: hierarchical reinforcement learning and task decomposition in humans." Computational and robotic models of the hierarchical organization of behavior: 271-291.

Durstewitz, D., N. M. Vittoz, S. B. Floresco and J. K. Seamans (2010). "Abrupt transitions between prefrontal neural ensemble states accompany behavioral transitions during rule learning." Neuron **66**(3): 438-448.

Duvelle, E., R. M. Grieves, A. Liu, S. Jedidi-Ayoub, J. Holeniewska, A. Harris, N. Nyberg, F. Donnarumma, J. M. Lefort, K. J. Jeffery, C. Summerfield, G. Pezzulo and H. J. Spiers (2021). "Hippocampal place cells encode global location but not connectivity in a complex space." Curr Biol **31**(6): 1221-1233 e1229.

Ego-Stengel, V. and M. A. Wilson (2010). "Disruption of ripple-associated hippocampal activity during rest impairs spatial learning in the rat." Hippocampus **20**(1): 1-10.

Eichenbaum, H. (1993). Memory, amnesia, and the hippocampal system, MIT press.

Ekstrom, A. D., M. J. Kahana, J. B. Caplan, T. A. Fields, E. A. Isham, E. L. Newman and I. Fried (2003). "Cellular networks underlying human spatial navigation." Nature **425**(6954): 184-188.

Espinosa, G., G. E. Wink, A. T. Lai, D. A. Dombeck and M. A. MacIver (2022). "Achieving mouse-level strategic evasion performance using real-time computational planning." arXiv preprint arXiv:2211.02700.

Fenton, A. A., H. Y. Kao, S. A. Neymotin, A. Olypher, Y. Vayntrub, W. W. Lytton and N. Ludvig (2008). "Unmasking the CA1 ensemble place code by exposures to small and large environments: more place cells and multiple, irregularly arranged, and expanded place fields in the larger space." J Neurosci **28**(44): 11250-11262.

Ferenczi, E. A., K. A. Zalocusky, C. Liston, L. Grosenick, M. R. Warden, D. Amatya, K. Katovich, H. Mehta, B. Patenaude, C. Ramakrishnan, P. Kalanithi, A. Etkin, B. Knutson, G. H. Glover and K. Deisseroth (2016). "Prefrontal cortical regulation of brainwide circuit dynamics and reward-related behavior." Science **351**(6268): aac9698.

Fernandez-Ruiz, A., A. Oliva, E. Fermino de Oliveira, F. Rocha-Almeida, D. Tingley and G. Buzsaki (2019). "Long-duration hippocampal sharp wave ripples improve memory." Science **364**(6445): 1082-1086.

Fortin, N. J., K. L. Agster and H. B. Eichenbaum (2002). "Critical role of the hippocampus in memory for sequences of events." Nat Neurosci **5**(5): 458-462.

Foster, D. J. (2017). "Replay Comes of Age." Annu Rev Neurosci **40**: 581-602.

Friedman, N. P. and T. W. Robbins (2022). "The role of prefrontal cortex in cognitive control and executive function." Neuropsychopharmacology **47**(1): 72-89.

Fujii, N. and A. M. Graybiel (2003). "Representation of action sequence boundaries by macaque prefrontal cortical neurons." Science **301**(5637): 1246-1249.

Gahnstrom, C. J. and H. J. Spiers (2020). "Striatal and hippocampal contributions to flexible navigation in rats and humans." Brain Neurosci Adv **4**: 2398212820979772.

Gardner, M. P. H., G. Schoenbaum and S. J. Gershman (2018). "Rethinking dopamine as generalized prediction error." Proc Biol Sci **285**(1891).

Gardner, R. S., P. E. Gold and D. L. Korol (2020). "Inactivation of the striatum in aged rats rescues their ability to learn a hippocampus-sensitive spatial navigation task." Neurobiology of learning and memory **172**: 107231.

Geerts, J. P., S. J. Gershman, N. Burgess and K. L. Stachenfeld (2023). "A probabilistic successor representation for context-dependent learning." Psychological Review.

George, A. E., J. J. Stout and A. L. Griffin (2023). "Pausing and reorienting behaviors enhance the performance of a spatial working memory task." Behav Brain Res **446**: 114410.

Gerfen, C. R. (1984). "The neostriatal mosaic: compartmentalization of corticostriatal input and striatonigral output systems." Nature **311**(5985): 461-464.

Gershman, S. J. (2018). "The Successor Representation: Its Computational Logic and Neural Substrates." J Neurosci **38**(33): 7193-7200.

Gillespie, A. K., D. A. Astudillo Maya, E. L. Denovellis, D. F. Liu, D. B. Kastner, M. E. Coulter, D. K. Roumis, U. T. Eden and L. M. Frank (2021). "Hippocampal replay reflects specific past experiences rather than a plan for subsequent choice." Neuron **109**(19): 3149-3163 e3146.

Glascher, J., N. Daw, P. Dayan and J. P. O'Doherty (2010). "States versus rewards: dissociable neural prediction error signals underlying model-based and model-free reinforcement learning." Neuron **66**(4): 585-595.

Gomperts, S. N., F. Kloosterman and M. A. Wilson (2015). "VTA neurons coordinate with the hippocampal reactivation of spatial experience." Elife **4**.

Graybiel, A. M. (1998). "The basal ganglia and chunking of action repertoires." Neurobiol Learn Mem **70**(1-2): 119-136.

Graybiel, A. M. (2008). "Habits, rituals, and the evaluative brain." Annu Rev Neurosci **31**: 359-387.

Gridchyn, I., P. Schoenenberger, J. O'Neill and J. Csicsvari (2020). "Assembly-Specific Disruption of Hippocampal Replay Leads to Selective Memory Deficit." Neuron **106**(2): 291-300 e296.

Groenewegen, H. J., Y. Galis-de Graaf and W. J. Smeets (1999). "Integration and segregation of limbic cortico-striatal loops at the thalamic level: an experimental tracing study in rats." J Chem Neuroanat **16**(3): 167-185.

Gupta, A. S., M. A. van der Meer, D. S. Touretzky and A. D. Redish (2010). "Hippocampal replay is not a simple function of experience." Neuron **65**(5): 695-705.

Gupta, A. S., M. A. van der Meer, D. S. Touretzky and A. D. Redish (2012). "Segmentation of spatial experience by hippocampal theta sequences." Nat Neurosci **15**(7): 1032-1039.

Hamrick, J. B., A. L. Friesen, F. Behbahani, A. Guez, F. Viola, S. Witherspoon, T. Anthony, L. Buesing, P. Veličković and T. Weber (2020). "On the role of planning in model-based deep reinforcement learning." arXiv preprint arXiv:2011.04021.

Hassabis, D., D. Kumaran and E. A. Maguire (2007). "Using imagination to understand the neural basis of episodic memory." J Neurosci **27**(52): 14365-14374.

Hasz, B. M. and A. D. Redish (2018). "Deliberation and Procedural Automation on a Two-Step Task for Rats." Front Integr Neurosci **12**: 30.

Hasz, B. M. and A. D. Redish (2020). "Dorsomedial prefrontal cortex and hippocampus represent strategic context even while simultaneously changing representation throughout a task session." Neurobiology of Learning and Memory **171**: 107215.

Heidbreder, C. A. and H. J. Groenewegen (2003). "The medial prefrontal cortex in the rat: evidence for a dorso-ventral distinction based upon functional and anatomical characteristics." Neurosci Biobehav Rev **27**(6): 555-579.

Heilbronner, S. R., J. Rodriguez-Romaguera, G. J. Quirk, H. J. Groenewegen and S. N. Haber (2016). "Circuit-Based Corticostriatal Homologies Between Rat and Primate." Biol Psychiatry **80**(7): 509-521.

Hikosaka, O., H. Nakahara, M. K. Rand, K. Sakai, X. Lu, K. Nakamura, S. Miyachi and K. Doya (1999). "Parallel neural networks for learning sequential procedures." Trends Neurosci **22**(10): 464-471.

Hitchcott, P. K., J. J. Quinn and J. R. Taylor (2007). "Bidirectional modulation of goal-directed actions by prefrontal cortical dopamine." Cereb Cortex **17**(12): 2820-2827.

Hok, V., E. Save, P. P. Lenck-Santini and B. Poucet (2005). "Coding for spatial goals in the prelimbic/infralimbic area of the rat frontal cortex." Proc Natl Acad Sci U S A **102**(12): 4602-4607.

Holroyd, C. B., J. J. F. Ribas-Fernandes, D. Shahnazian, M. Silvetti and T. Verguts (2018). "Human midcingulate cortex encodes distributed representations of task progress." Proc Natl Acad Sci U S A **115**(25): 6398-6403.

Hoover, W. B. and R. P. Vertes (2007). "Anatomical analysis of afferent projections to the medial prefrontal cortex in the rat." Brain Struct Funct **212**(2): 149-179.

Hull, C. L. (1943). "Principles of behavior: an introduction to behavior theory."

Hunnicutt, B. J., B. C. Jongbloets, W. T. Birdsong, K. J. Gertz, H. Zhong and T. Mao (2016). "A comprehensive excitatory input map of the striatum reveals novel functional organization." Elife **5**.

Hunt, L., N. Daw, P. Kaanders, M. MacIver, U. Mugan, E. Procyk, A. Redish, E. Russo, J. Scholl and K. Stachenfeld (2021). "Formalizing planning and information search in naturalistic decision-making." Nature neuroscience **24**(8): 1051-1064.

Hurley, K. M., H. Herbert, M. M. Moga and C. B. Saper (1991). "Efferent projections of the infralimbic cortex of the rat." J Comp Neurol **308**(2): 249-276.

Huys, Q. J., N. Eshel, E. O'Nions, L. Sheridan, P. Dayan and J. P. Roiser (2012). "Bonsai trees in your head: how the pavlovian system sculpts goal-directed choices by pruning decision trees." PLoS Comput Biol **8**(3): e1002410.

Ito, M. and K. Doya (2011). "Multiple representations and algorithms for reinforcement learning in the cortico-basal ganglia circuit." Curr Opin Neurobiol **21**(3): 368-373.

Jadhav, S. P., C. Kemere, P. W. German and L. M. Frank (2012). "Awake hippocampal sharp-wave ripples support spatial memory." Science **336**(6087): 1454-1458.

Jeffery, K. J. and M. I. Anderson (2003). "Dissociation of the geometric and contextual influences on place cells." Hippocampus **13**(7): 868-872.

Jin, X. and R. M. Costa (2010). "Start/stop signals emerge in nigrostriatal circuits during sequence learning." Nature **466**(7305): 457-462.

Jin, X., F. Tecuapetla and R. M. Costa (2014). "Basal ganglia subcircuits distinctively encode the parsing and concatenation of action sequences." Nat Neurosci **17**(3): 423-430.

Joel, D., Y. Niv and E. Ruppin (2002). "Actor-critic models of the basal ganglia: new anatomical and computational perspectives." Neural Netw **15**(4-6): 535-547.

Jog, M. S., Y. Kubota, C. I. Connolly, V. Hillegaart and A. M. Graybiel (1999). "Building neural representations of habits." Science **286**(5445): 1745-1749.

Johnson, A. and A. D. Redish (2007). "Neural ensembles in CA3 transiently encode paths forward of the animal at a decision point." J Neurosci **27**(45): 12176-12189.

Ju, H. and D. S. Bassett (2020). "Dynamic representations in networked neural systems." Nat Neurosci **23**(8): 908-917.

Judak, L., B. Chiovini, G. Juhasz, D. Palfi, Z. Mezriczky, Z. Szadai, G. Katona, B. Szmola, K. Ocsai, B. Martinecz, A. Mihaly, A. Denes, B. Kerekes, A. Szepesi, G. Szalay, I. Ulbert, Z. Mucsi, B. Roska and B. Rozsa (2022). "Sharp-wave ripple doublets induce complex dendritic spikes in parvalbumin interneurons in vivo." Nat Commun **13**(1): 6715.

Karlsson, M. P., D. G. Tervo and A. Y. Karpova (2012). "Network resets in medial prefrontal cortex mark the onset of behavioral uncertainty." Science **338**(6103): 135-139.

Karuza, E. A., S. L. Thompson-Schill and D. S. Bassett (2016). "Local Patterns to Global Architectures: Influences of Network Topology on Human Learning." Trends Cogn Sci **20**(8): 629-640.

Kay, K., J. E. Chung, M. Sosa, J. S. Schor, M. P. Karlsson, M. C. Larkin, D. F. Liu and L. M. Frank (2020). "Constant Sub-second Cycling between Representations of Possible Futures in the Hippocampus." Cell **180**(3): 552-567 e525.

Kesner, R. P. and J. C. Churchwell (2011). "An analysis of rat prefrontal cortex in mediating executive function." Neurobiol Learn Mem **96**(3): 417-431.

Khamassi, M., A. B. Mulder, E. Tabuchi, V. Douchamps and S. I. Wiener (2008). "Anticipatory reward signals in ventral striatal neurons of behaving rats." Eur J Neurosci **28**(9): 1849-1866.

Kidder, K. S., J. T. Miles, P. M. Baker, V. I. Hones, D. H. Gire and S. J. Y. Mizumori (2021). "A selective role for the mPFC during choice and deliberation, but not spatial memory retention over short delays." Hippocampus **31**(7): 690-700.

Killcross, S. and E. Coutureau (2003). "Coordination of actions and habits in the medial prefrontal cortex of rats." Cereb Cortex **13**(4): 400-408.

Knierim, J. J., H. S. Kudrimoti and B. L. McNaughton (1995). "Place cells, head direction cells, and the learning of landmark stability." J Neurosci **15**(3 Pt 1): 1648-1659.

Knutson, B., J. Taylor, M. Kaufman, R. Peterson and G. Glover (2005). "Distributed neural representation of expected value." J Neurosci **25**(19): 4806-4812.

Krausz, T. A., A. E. Comrie, L. M. Frank, N. D. Daw and J. D. Berke (2023). "Dual credit assignment processes underlie dopamine signals in a complex spatial environment." bioRxiv.

Kwak, S. and M. W. Jung (2019). "Distinct roles of striatal direct and indirect pathways in value-based decision making." Elife **8**: e46050.

Lally, N., Q. J. M. Huys, N. Eshel, P. Faulkner, P. Dayan and J. P. Roiser (2017). "The Neural Basis of Aversive Pavlovian Guidance during Planning." J Neurosci **37**(42): 10215-10229.

Laubach, M., L. M. Amarante, K. Swanson and S. R. White (2018). "What, If Anything, Is Rodent Prefrontal Cortex?" eNeuro **5**(5).

Lavoie, A. M. and S. J. Mizumori (1994). "Spatial, movement- and reward-sensitive discharge by medial ventral striatum neurons of rats." Brain Res **638**(1-2): 157-168.

Lee, J. Q., D. O. LeDuke, K. Chua, R. J. McDonald and R. J. Sutherland (2018). "Relocating cued goals induces population remapping in CA1 related to memory performance in a two-platform water task in rats." Hippocampus **28**(6): 431-440.

Leutgeb, S., J. K. Leutgeb, C. A. Barnes, E. I. Moser, B. L. McNaughton and M. B. Moser (2005). "Independent codes for spatial and episodic memory in hippocampal neuronal ensembles." Science **309**(5734): 619-623.

Lever, C., T. Wills, F. Cacucci, N. Burgess and J. O'Keefe (2002). "Long-term plasticity in hippocampal place-cell representation of environmental geometry." Nature **416**(6876): 90-94.

Lisman, J. (2015). "The challenge of understanding the brain: where we stand in 2015." Neuron **86**(4): 864-882.

Lisman, J. and A. D. Redish (2009). "Prediction, sequences and the hippocampus." Philosophical Transactions of the Royal Society B: Biological Sciences **364**(1521): 1193-1201.

Maguire, E. A., R. Nannery and H. J. Spiers (2006). "Navigation around London by a taxi driver with bilateral hippocampal lesions." Brain **129**(Pt 11): 2894-2907.

McLaughlin, A. E., G. W. Diehl and A. D. Redish (2021). "Potential roles of the rodent medial prefrontal cortex in conflict resolution between multiple decision-making systems." Int Rev Neurobiol **158**: 249-281.

McNamara, C. G., A. Tejero-Cantero, S. Trouche, N. Campo-Urriza and D. Dupret (2014). "Dopaminergic neurons promote hippocampal reactivation and spatial memory persistence." Nat Neurosci **17**(12): 1658-1660.

Miller, K. J., M. M. Botvinick and C. D. Brody (2017). "Dorsal hippocampus contributes to model-based planning." Nat Neurosci **20**(9): 1269-1276.

Miller, K. J., E. A. Ludvig, G. Pezzulo and A. Shenhav (2018). Realigning models of habitual and goal-directed decision-making. Goal-directed decision making, Elsevier**:** 407-428.

Mnih, V., K. Kavukcuoglu, D. Silver, A. A. Rusu, J. Veness, M. G. Bellemare, A. Graves, M. Riedmiller, A. K. Fidjeland and G. Ostrovski (2015). "Human-level control through deep reinforcement learning." nature **518**(7540): 529-533.

Mobbs, D., D. B. Headley, W. Ding and P. Dayan (2020). "Space, time, and fear: survival computations along defensive circuits." Trends in cognitive sciences **24**(3): 228-241.

Morris, R. G., P. Garrud, J. N. Rawlins and J. O'Keefe (1982). "Place navigation impaired in rats with hippocampal lesions." Nature **297**(5868): 681-683.

Mugan, U., S. L. Hoffman, P. J. Cunnigham, P. S. Regier, S. Amemiya and A. D. Redish (2022). Environmental complexity modulates the arbitration between deliberative and habitual decision-making. Computational and Systems Neuroscience (COSYNE), Lisbon, Portugal.

Mugan, U. and M. A. MacIver (2020). "Spatial planning with long visual range benefits escape from visual predators in complex naturalistic environments." Nat Commun **11**(1): 3057.

Mukherjee, A. and P. Caroni (2018). "Infralimbic cortex is required for learning alternatives to prelimbic promoted associations through reciprocal connectivity." Nat Commun **9**(1): 2727.

Newman, M. E. J., A.-L. s. Barabási and D. J. Watts (2006). The structure and dynamics of networks. Princeton, Princeton University Press.

Niv, Y. (2009). "Reinforcement learning in the brain." Journal of Mathematical Psychology **53**(3): 139-154.

O'Keefe, J. and J. Dostrovsky (1971). "The hippocampus as a spatial map. Preliminary evidence from unit activity in the freely-moving rat." Brain Res **34**(1): 171-175.

O'Keefe, J. and L. Nadel (1978). The hippocampus as a cognitive map. Oxford

New York, Clarendon Press ;

Oxford University Press.

O'Keefe, J., L. Nadel, S. Keightley and D. Kill (1975). "Fornix lesions selectively abolish place learning in the rat." Exp Neurol **48**(1): 152-166.

Olafsdottir, H. F., C. Barry, A. B. Saleem, D. Hassabis and H. J. Spiers (2015). "Hippocampal place cells construct reward related sequences through unexplored space." Elife **4**: e06063.

Padilla-Coreano, N., S. Canetta, R. M. Mikofsky, E. Alway, J. Passecker, M. V. Myroshnychenko, A. L. Garcia-Garcia, R. Warren, E. Teboul, D. R. Blackman, M. P. Morton, S. Hupalo, K. M. Tye, C. Kellendonk, D. A. Kupferschmidt and J. A. Gordon (2019). "Hippocampal-Prefrontal Theta Transmission Regulates Avoidance Behavior." Neuron **104**(3): 601-610 e604.

Papale, A. E., M. C. Zielinski, L. M. Frank, S. P. Jadhav and A. D. Redish (2016). "Interplay between Hippocampal Sharp-Wave-Ripple Events and Vicarious Trial and Error Behaviors in Decision Making." Neuron **92**(5): 975-982.

Peters, A. J., J. M. J. Fabre, N. A. Steinmetz, K. D. Harris and M. Carandini (2021). "Striatal activity topographically reflects cortical activity." Nature **591**(7850): 420-425.

Pfeiffer, B. E. and D. J. Foster (2013). "Hippocampal place-cell sequences depict future paths to remembered goals." Nature **497**(7447): 74-79.

Pooters, T., I. Gantois, B. Vermaercke and R. D'Hooge (2016). "Inability to acquire spatial information and deploy spatial search strategies in mice with lesions in dorsomedial striatum." Behav Brain Res **298**(Pt B): 134-141.

Powell, N. J. and A. D. Redish (2014). "Complex neural codes in rat prelimbic cortex are stable across days on a spatial decision task." Front Behav Neurosci **8**: 120.

Powell, N. J. and A. D. Redish (2016). "Representational changes of latent strategies in rat medial prefrontal cortex precede changes in behaviour." Nat Commun **7**: 12830.

Preston, A. R. and H. Eichenbaum (2013). "Interplay of hippocampus and prefrontal cortex in memory." Curr Biol **23**(17): R764-773.

Rawlins, J. (1985). "Associations across time: The hippocampus as a temporary memory store." Behavioral and Brain Sciences **8**(3): 479-497.

Redish, A. D. (1999). Beyond the cognitive map: from place cells to episodic memory, MIT press.

Redish, A. D. (2013). The mind within the brain : how we make decisions and how those decisions go wrong. Oxford, Oxford University Press.

Redish, A. D. (2016). "Vicarious trial and error." Nat Rev Neurosci **17**(3): 147-159.

Regier, P. S., S. Amemiya and A. D. Redish (2015). "Hippocampus and subregions of the dorsal striatum respond differently to a behavioral strategy change on a spatial navigation task." J Neurophysiol **114**(3): 1399-1416.

Rich, E. L. and M. Shapiro (2009). "Rat prefrontal cortical neurons selectively code strategy switches." Journal of Neuroscience **29**(22): 7208-7219.

Richards, B. A., T. P. Lillicrap, P. Beaudoin, Y. Bengio, R. Bogacz, A. Christensen, C. Clopath, R. P. Costa, A. de Berker and S. Ganguli (2019). "A deep learning framework for neuroscience." Nature neuroscience **22**(11): 1761-1770.

Robbins, T. and R. M. Costa (2017). "Habits." Current biology **27**(22): R1200-R1206.

Rushworth, M. F. and T. E. Behrens (2008). "Choice, uncertainty and value in prefrontal and cingulate cortex." Nature neuroscience **11**(4): 389-397.

Schapiro, A. C., N. B. Turk-Browne, M. M. Botvinick and K. A. Norman (2017). "Complementary learning systems within the hippocampus: a neural network modelling approach to reconciling episodic memory with statistical learning." Philos Trans R Soc Lond B Biol Sci **372**(1711).

Schmidt, B., A. A. Duin and A. D. Redish (2019). "Disrupting the medial prefrontal cortex alters hippocampal sequences during deliberative decision making." J Neurophysiol **121**(6): 1981-2000.

Schmitzer-Torbert, N. and A. D. Redish (2002). "Development of path stereotypy in a single day in rats on a multiple-T maze." Arch Ital Biol **140**(4): 295-301.

Schultz, W., P. Dayan and P. R. Montague (1997). "A neural substrate of prediction and reward." Science **275**(5306): 1593-1599.

Shamash, P., S. Lee, A. M. Saxe and T. Branco (2023). "Mice identify subgoal locations through an action-driven mapping process." Neuron.

Shamash, P., S. F. Olesen, P. Iordanidou, D. Campagner, N. Banerjee and T. Branco (2021). "Mice learn multi-step routes by memorizing subgoal locations." Nat Neurosci **24**(9): 1270-1279.

Shaw, C. L., G. D. R. Watson, H. L. Hallock, K. M. Cline and A. L. Griffin (2013). "The role of the medial prefrontal cortex in the acquisition, retention, and reversal of a tactile visuospatial conditional discrimination task." Behav Brain Res **236**(1): 94-101.

Siegel, M., T. H. Donner and A. K. Engel (2012). "Spectral fingerprints of large-scale neuronal interactions." Nat Rev Neurosci **13**(2): 121-134.

Silver, D., J. Schrittwieser, K. Simonyan, I. Antonoglou, A. Huang, A. Guez, T. Hubert, L. Baker, M. Lai and A. Bolton (2017). "Mastering the game of go without human knowledge." nature **550**(7676): 354-359.

Singh, S. P. (1992). Reinforcement learning with a hierarchy of abstract models. Proceedings of the National Conference on Artificial Intelligence, Citeseer.

Smith, K. S. and A. M. Graybiel (2013). "A dual operator view of habitual behavior reflecting cortical and striatal dynamics." Neuron **79**(2): 361-374.

Smith, K. S. and A. M. Graybiel (2013). "Using optogenetics to study habits." Brain Res **1511**: 102-114.

Smith, K. S. and A. M. Graybiel (2016). "Habit formation coincides with shifts in reinforcement representations in the sensorimotor striatum." J Neurophysiol **115**(3): 1487-1498.

Smith, K. S., A. Virkud, K. Deisseroth and A. M. Graybiel (2012). "Reversible online control of habitual behavior by optogenetic perturbation of medial prefrontal cortex." Proc Natl Acad Sci U S A **109**(46): 18932-18937.

Solstad, T., E. I. Moser and G. T. Einevoll (2006). "From grid cells to place cells: a mathematical model." Hippocampus **16**(12): 1026-1031.

Spellman, T., M. Rigotti, S. E. Ahmari, S. Fusi, J. A. Gogos and J. A. Gordon (2015). "Hippocampal-prefrontal input supports spatial encoding in working memory." Nature **522**(7556): 309-314.

Stachenfeld, K. L., M. M. Botvinick and S. J. Gershman (2017). "The hippocampus as a predictive map." Nat Neurosci **20**(11): 1643-1653.

Sutton, R. S. and A. G. Barto (2018). Reinforcement learning : an introduction. Cambridge, Massachusetts, The MIT Press.

Thorn, C. A., H. Atallah, M. Howe and A. M. Graybiel (2010). "Differential dynamics of activity changes in dorsolateral and dorsomedial striatal loops during learning." Neuron **66**(5): 781-795.

Tolman, E. C. (1939). "Prediction of vicarious trial and error by means of the schematic sowbug." Psychological Review **46**(4): 318.

Tolman, E. C. (1948). "Cognitive maps in rats and men." Psychol Rev **55**(4): 189-208.

Trucco, E. (1956). "A note on the information content of graphs." The bulletin of mathematical biophysics **18**(2): 129-135.

Ulanovsky, N. and C. F. Moss (2007). "Hippocampal cellular and network activity in freely moving echolocating bats." Nat Neurosci **10**(2): 224-233.

Uylings, H. B., H. J. Groenewegen and B. Kolb (2003). "Do rats have a prefrontal cortex?" Behav Brain Res **146**(1-2): 3-17.

van der Meer, M. A., A. Johnson, N. C. Schmitzer-Torbert and A. D. Redish (2010). "Triple dissociation of information processing in dorsal striatum, ventral striatum, and hippocampus on a learned spatial decision task." Neuron **67**(1): 25-32.

van der Meer, M. A., T. Kalenscher, C. S. Lansink, C. M. Pennartz, J. D. Berke and A. D. Redish (2010). "Integrating early results on ventral striatal gamma oscillations in the rat." Front Neurosci **4**: 300.

van der Meer, M. A. and A. D. Redish (2009). "Covert Expectation-of-Reward in Rat Ventral Striatum at Decision Points." Front Integr Neurosci **3**: 1.

van der Meer, M. A. and A. D. Redish (2010). "Expectancies in decision making, reinforcement learning, and ventral striatum." Front Neurosci **4**: 6.

van der Meer, M. A. and A. D. Redish (2011). "Theta phase precession in rat ventral striatum links place and reward information." J Neurosci **31**(8): 2843-2854.

van der Meer, M. A. and A. D. Redish (2011). "Ventral striatum: a critical look at models of learning and evaluation." Curr Opin Neurobiol **21**(3): 387-392.

Vidal-Gonzalez, I., B. Vidal-Gonzalez, S. L. Rauch and G. J. Quirk (2006). "Microstimulation reveals opposing influences of prelimbic and infralimbic cortex on the expression of conditioned fear." Learn Mem **13**(6): 728-733.

Voorn, P., L. J. Vanderschuren, H. J. Groenewegen, T. W. Robbins and C. M. Pennartz (2004). "Putting a spin on the dorsal-ventral divide of the striatum." Trends Neurosci **27**(8): 468-474.

Whittington, J. C., T. H. Muller, S. Mark, G. Chen, C. Barry, N. Burgess and T. E. Behrens (2019). "The Tolman-Eichenbaum Machine: Unifying space and relational memory through generalisation in the hippocampal formation." bioRxiv: 770495.

Wikenheiser, A. M. and A. D. Redish (2015). "Decoding the cognitive map: ensemble hippocampal sequences and decision making." Curr Opin Neurobiol **32**: 8-15.

Wikenheiser, A. M. and A. D. Redish (2015). "Hippocampal theta sequences reflect current goals." Nat Neurosci **18**(2): 289-294.

Wilson, R. C., E. Bonawitz, V. D. Costa and R. B. Ebitz (2021). "Balancing exploration and exploitation with information and randomization." Current opinion in behavioral sciences **38**: 49-56.

Wu, X. and D. J. Foster (2014). "Hippocampal replay captures the unique topological structure of a novel environment." J Neurosci **34**(19): 6459-6469.

Yamamoto, J. and S. Tonegawa (2017). "Direct Medial Entorhinal Cortex Input to Hippocampal CA1 Is Crucial for Extended Quiet Awake Replay." Neuron **96**(1): 217-227 e214.

Yin, H. H. and B. J. Knowlton (2006). "The role of the basal ganglia in habit formation." Nat Rev Neurosci **7**(6): 464-476.

Yin, H. H., B. J. Knowlton and B. W. Balleine (2004). "Lesions of dorsolateral striatum preserve outcome expectancy but disrupt habit formation in instrumental learning." Eur J Neurosci **19**(1): 181-189.

Yin, H. H., B. J. Knowlton and B. W. Balleine (2006). "Inactivation of dorsolateral striatum enhances sensitivity to changes in the action-outcome contingency in instrumental conditioning." Behav Brain Res **166**(2): 189-196.

Zhang, H., P. D. Rich, A. K. Lee and T. O. Sharpee (2023). "Hippocampal spatial representations exhibit a hyperbolic geometry that expands with experience." Nat Neurosci **26**(1): 131-139.

Zhang, K., I. Ginzburg, B. L. McNaughton and T. J. Sejnowski (1998). "Interpreting neuronal population activity by reconstruction: unified framework with application to hippocampal place cells." J Neurophysiol **79**(2): 1017-1044.

---

## Figure 1: Different algorithms of decision-making: Predicted behavioral and neural manifestations.

First column indicates a simple example of how each algorithm would evaluate states and choose actions, the second column indicates behavioral predictions based on the corresponding decision-making algorithm, and the third column indicates neural predictions based on the corresponding decision-making algorithm. There is no explicit relationship between the simple examples in column one, and the illustrated experimental paradigms in column two. (A1) Model-based algorithms represent all of the states and how they are connected to each other. Actions are chosen based on a search. After contingency change, the value expectations change and are broadcasted to all other connected states. (A2) Paths are variable as the animal searches through its mental space. (A3) Dorsal hippocampal place cells alternately sweeping to the next goal with each theta cycle may underlie the planning process. (B1) Model-free decisions learn optimal actions at each state, but do not represent states that are not local. After contingency change, these reassignments have to be relearned. (B2) The fast and reliable table-lookup computations create stereotyped paths. (B3, left) Illustration of DLS cells developing representations at important decision-points, and schematic of classic task-bracketing that do not include multiple turns. (B3, right) Summed DLS medium spiny neuron activity on an LRA task with multiple internal turns (T1, T2, T3, T4); note the increased activity along the complex central track (turns 1-4), but not on the simpler return journey (after the two food reward sides (F1, F2), unlike what would be predicted with classical task-bracketing (B3, right). Panel from van der Meer, Johnson et al. (2010). (C1) Successor representation theory suggests that repeated paths (here states 0 to 1 to 4) can become represented as a single entity (S014; green overlay) so that planning can occur over them. (C2) Behaviorally, repeated trajectories should be maintained even when contingencies are changed, which suggests that any stereotyped paths that develop should be maintained throughout. (C3) Neurophysiologically, current successor representation theories suggest that dorsal hippocampal place cells (each colored blob) should encode repeated paths which is dependent on task structure and task reward structure. (D1) Chunked action theories suggest that decisions are made over the entire journey to the next goal. Planning over chunked action theories suggest that decisions are made based on imagined outcome evaluations. Planning takes place over connected reliably valuable action chunks. (D2) Behaviorally, paths should be stereotyped, but should change between contingencies. Red highlight shows the original path before a contingency change. (D3) Neurophysiologically, planning over cached action sequences theory suggests that task-bracketing (increased firing at the start of an action bout) should encode future outcomes. Adapted from Cunningham, Regier et al. (2021). (E1) Habit over chunked actions suggests that with a contingency change only the immediate value of the chunk is updated, and the value similar to model-free methods, is thought to backpropagate to the starting abstract state. Therefore, cached action theories suggest that decisions are made based on learned actions associated with situations. (E2) Behaviorally, this theory predicts that each contingency will have its own set of stereotyped actions. (E3) Neurophysiologically, habit over cached action sequences theory suggests that task-bracketing information should be about current and past situation, not future ones, unlike (D3). Adapted from Cunningham, Regier et al. (2021).

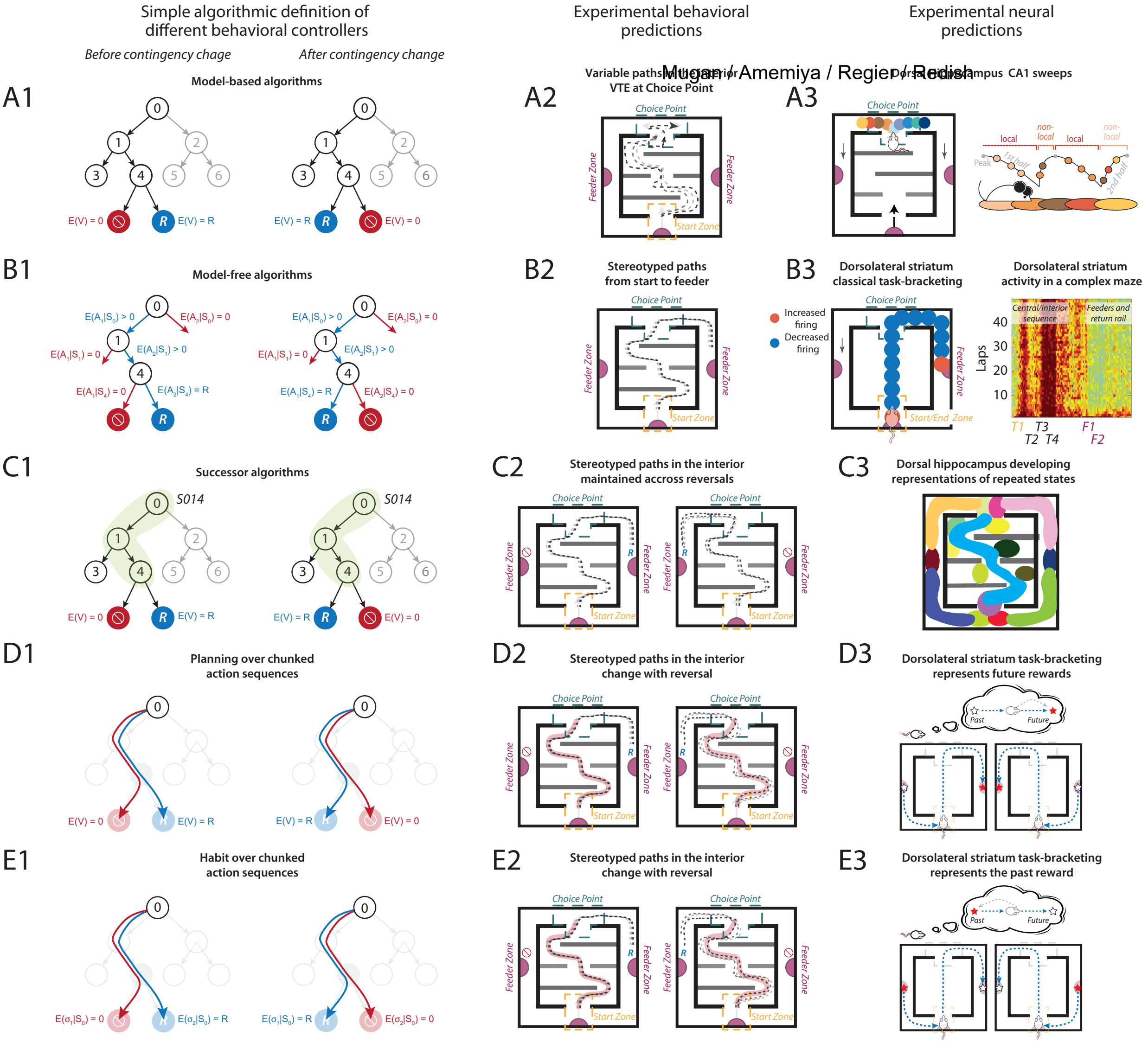
Simple algorithmic definition of different behavioral controllers
Experimental behavioral predictions
Experimental neural predictions
Before contingency chage
After contingency change
A1
Model-based algorithms
E(V) = 0
E(V) = R
A2
Variable paths in the interior
VTE at Choice Point
Choice Point
Feeder Zone
Start Zone
A3
Dorsal hippocampus CA1 sweeps
local
non-local
Peak
1st half
2nd half
B1
Model-free algorithms
B2
Stereotyped paths from start to feeder
B3
Dorsolateral striatum classical task-bracketing
Increased firing
Decreased firing
Start/End Zone
Dorsolateral striatum activity in a complex maze
Central/interior sequence
Feeders and return rail
Laps
T1 T3 T2 T4 F1 F2
C1
Successor algorithms
S014
C2
Stereotyped paths in the interior maintained accross reversals
C3
Dorsal hippocampus developing representations of repeated states
D1
Planning over chunked action sequences
D2
Stereotyped paths in the interior change with reversal
D3
Dorsolateral striatum task-bracketing represents future rewards
Past
Future
E1
Habit over chunked action sequences
E2
Stereotyped paths in the interior change with reversal
E3
Dorsolateral striatum task-bracketing represents the past reward

## Figure 2: Behavioral and neural manifestations of decision-making in environments with varying levels of complexity.

(A) The LRA task is a spatial maze with a choice point where rats have to make a decision to go either left or right for a food reward. Contingencies are presented in blocks and a contingency change occurs roughly half-way through the session (~15 mins). On the right is an example behavioral data from a single session. (B) Environmental complexity with respect to central path structure. (C) Median overall path variance — proxy for path variability — for environments with different complexities. (D) Two example laps depicting VTE and non-VTE paths through the high-cost choice point. VTE was identified by a log threshold value (value: 4) on the integrated angular velocity across a pass through the choice point. (E) Correlation between central path complexity and proportion of identified VTE laps in a given session. Shaded region: 95% CI. (F) Example low and high complexities environment overlaid with two example paths (blue and purple) through the internal central track. Notice that the blue and purple paths overlap more in low complexity environments. (G) Path stereotypy was calculated as the lap-by-lap reciprocal Euclidean distance. (G; Left column) Two example sessions showing path stereotypy in low and high complexity environments. Notice that path stereotypy develops throughout a stable contingency but decreases after a contingency switch and redevelops. (G; Right column) Average ($n_{Rats}$ = 11) path stereotypy through the central track aligned to switch times. Similar progression in path stereotypy can be seen as the examples. (H) Distribution of theta Asymmetry Index (AI) at the choice point in environments with low (cyan) and high (purple) path complexity ($n_{Rats}$ = 5; MWU $p < 10^{-4}$). (I) Box plot of 1st half and 2nd half of theta cycle durations at the choice point grouped according to path complexity (cyan: low, purple: high) ($n_{Rats}$ = 5; MWU $p_{1st\text{-}half} = 0.26$, $p_{2nd\text{-}half} < 10^{-5}$). (J) Distribution of the durations of SWRs in environments with low (cyan) and high (purple) path complexity ($n_{Rats}$ = 5; MWU $p < 10^{-4}$). (K) Proportion of SWRs occurring in isolation or as doublets or triplets grouped according to path complexity (cyan: low, purple: high). (L) Correlation of task-bracketing index and path complexity throughout the task (top panel: first contingency, bottom panel: second contingency). Grey shaded region indicates the 95% CI ($n_{Rats}$ = 6). For all analysis low path complexity sessions were identified as those below the 25$^{th}$ percentile, and high path complexity sessions were identified as those above the 75$^{th}$ percentile.

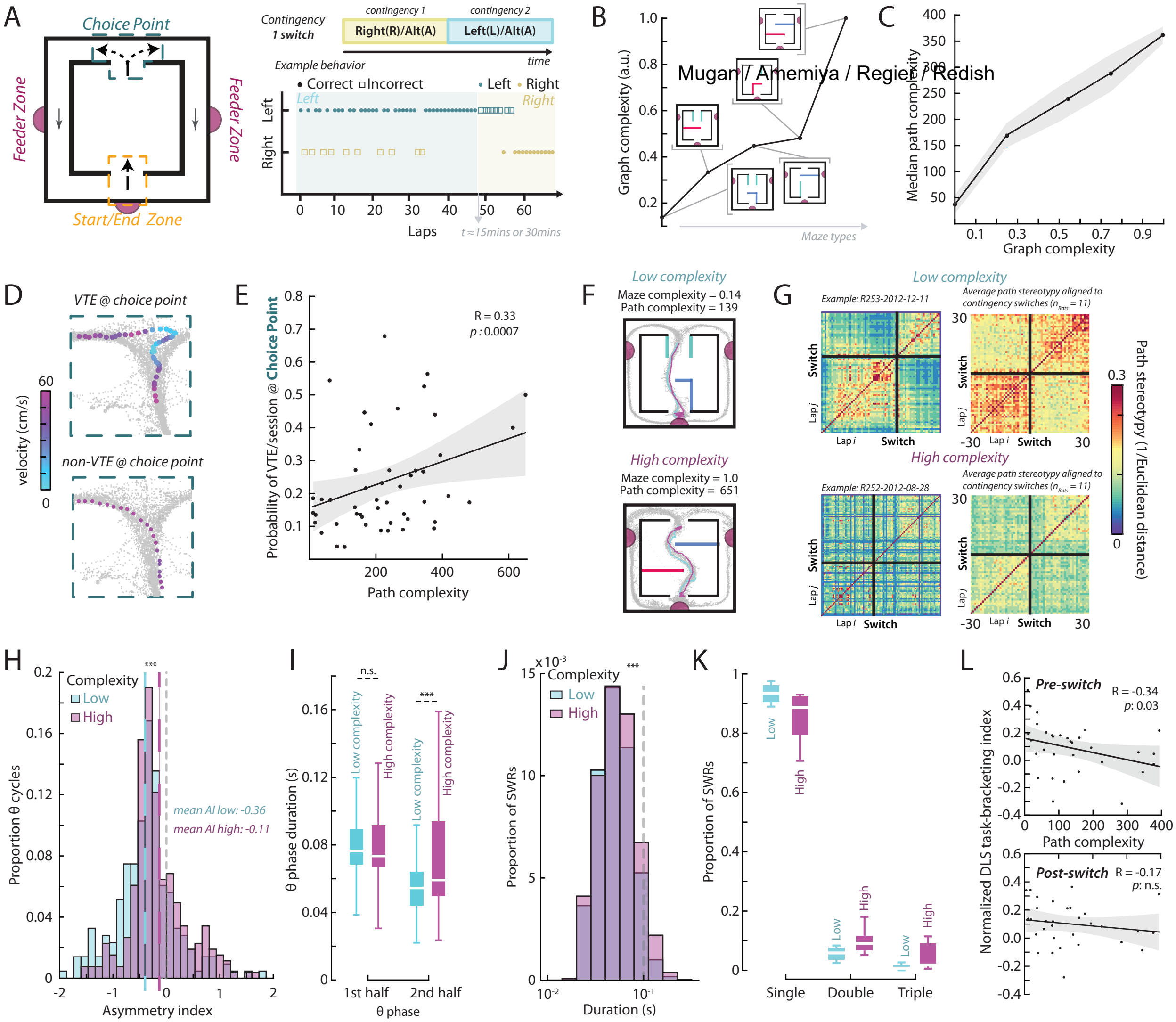